\documentclass[prl,aps,twocolumn,doublespace,floatfix,showpacs,superscriptaddress]{revtex4}
\usepackage{graphicx}

\begin{document}

\title{Universal spin-induced Time Reversal Symmetry breaking in two-dimensional electron gases with Rashba spin-orbit interaction}

\author{F.E. Meijer} \email{f.e.meijer@tnw.tudelft.nl}
\affiliation{Kavli Institute of NanoScience, Delft University of
Technology,  Lorentzweg 1, 2628 CJ Delft, The Netherlands}
\author{A.F. Morpurgo}
\affiliation{Kavli Institute of NanoScience, Delft University of
Technology,  Lorentzweg 1, 2628 CJ Delft, The Netherlands}
\author{T.M. Klapwijk}
\affiliation{Kavli Institute of NanoScience, Delft University of
Technology,  Lorentzweg 1, 2628 CJ Delft, The Netherlands}
\author{J. Nitta}
\altaffiliation[present address: ]{Department of Materials
Science, Tohoku University, Sendai 980-8579, Japan.}
\affiliation{NTT Basic Research Laboratories, NTT Corporation,
Atsugi-shi, Kanagawa 243-0198, Japan} \affiliation{CREST-Japan
Science and Technology Agency}

\date{\today}

\begin{abstract}
We have experimentally studied the spin-induced time reversal
symmetry (TRS) breaking as a function of the relative strength of
the Zeeman energy ($E_Z$) and the Rashba spin-orbit interaction
energy ($E_{SOI}$), in InGaAs-based 2D electron gases. We find
that the TRS breaking, and hence the associated dephasing time
$\tau_\phi(B)$, \textit{saturates} when $E_Z$ becomes comparable
to $E_{SOI}$. Moreover, we show that the spin-induced TRS breaking
mechanism is a \textit{universal function} of the ratio
$E_Z/E_{SOI}$, within the experimental accuracy.
\end{abstract}


\maketitle

The spin dynamics in solid state systems is commonly determined by
the competition between two energy scales; the Zeeman energy and
spin-orbit interaction (SOI) energy. If the Zeeman energy ($E_Z$)
is dominant, the spin is always aligned with the applied magnetic
field. In contrast, if the spin-orbit interaction is dominant, the
spin and orbital dynamics are coupled, and elastic scattering
therefore randomizes the spin precession axis. This results in a
finite spin relaxation time $\tau_s(0)$ \cite{DP}. Hence, the
"control parameter" for the spin dynamics in diffusive systems is
the ratio $E_Z/E_{SOI}$, where $E_{SOI} \equiv \hbar/\tau_s(0)$.
Consequently, many proposals and physical phenomena in the field
of spintronics depend on the ratio of these two energy
scales\cite{schwab,winkler,dots}.

An example where the spin dynamics, and therefore the ratio $E_Z /
E_{SOI}$, plays an important role is in phase-coherent transport:
quantum interference is qualitatively different for $E_Z / E_{SOI}
\ll 1$ and $E_Z / E_{SOI} \gg 1$. For $E_Z / E_{SOI} \rightarrow
\infty$ the spin dynamics does not depend on the orbital motion of
the electrons (Fig. 1a). The spin is a good quantum number and the
interference takes place within each spin-subband separately. For
$E_Z / E_{SOI} \ll 1$, the spin is not a conserved quantity, and
the spin randomly precesses during the orbital motion (Fig. 1b).
This leads to mixing of the spin subbands in the interference
process (resulting in weak-anti localization\cite{bergmann,ILP}).
Increasing the ratio $E_Z / E_{SOI}$ leads therefore to a
crossover between two conceptually different physical conditions.

In the limit $E_Z/E_{SOI} \ll 1$, it was recently shown
theoretically\cite{malshukov} and
experimentally\cite{meijer,minkov} that increasing the ratio
$E_Z/E_{SOI}$ from 0 to a finite value ($ \ll 1$), results in
dephasing of time-reversed paths, i.e. it induces Time Reversal
Symmetry (TRS) breaking. The effect of the interplay between
Zeeman and SOI on quantum interference is therefore quite similar
to a small perpendicular magnetic field (i.e. a magnetic flux);
they both introduce an upper time-scale for interference, which is
shorter than the inelastic scattering time\cite{time scale}. We
denote this upper time-scale due to the interplay between Zeeman
and SOI by $\tau_{\phi}(B_{\parallel})$.

\begin{figure}[b]
\begin{center}\leavevmode
\includegraphics[width=1\linewidth]{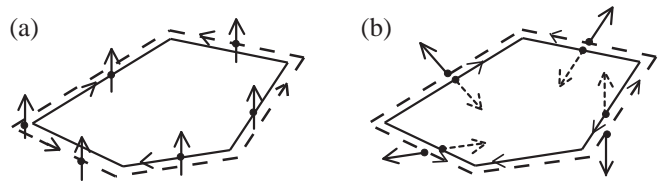}
\caption{Schematic illustration of the relevant time-reversed
trajectories contributing to weak-(anti)localization for the case
of negligible SOI (a; $E_Z/E_{SOI} \rightarrow \infty$) and strong
SOI (b; $E_{Z}/E_{SOI} \ll 1$). If SOI is negligible the spin
remains parallel to itself and interference takes place for each
spin-subband separately. If SOI is strong, the spin dynamics has
to be taken into account. This changes the nature of time-reversed
trajectories and makes them sensitive to the presence of a
magnetic field applied parallel to the plane of the
two-dimensional electron gas.} \label{B}
\end{center}
\end{figure}

In this paper we investigate experimentally the TRS breaking, due
to the competition between Zeeman coupling and Rashba
SOI\cite{rashba}, for the whole range of $E_Z/E_{SOI}$, i.e up to
$E_Z/E_{SOI} \gg 1$. We demonstrate that the spin-induced TRS
breaking, and hence the associated dephasing time $\tau_\phi
(B_{\parallel})$, \textit{saturates} when $E_Z/E_{SOI} \approx 1$,
i.e. when the spin becomes aligned with the external magnetic
field. The saturation value of the dephasing time
$\tau_{\phi}(B_{\parallel})$ is found to depend exclusively on the
spin relaxation time $\tau_s(0)$. Moreover, we show that the
quantity $\tau_s(0)/\tau_{\phi}(B_{\parallel})$ is a
\textit{universal function} of $E_Z/E_{SOI}$, i.e. it is
independent of any details of the quantum  well, such as the
electron density, elastic scattering time and Rashba spin-split
energy $\Delta$. All these conclusions are based on the detailed
quantitative analysis of the magnetoconductance as a function of
perpendicular and parallel magnetic field.

In our investigation we have used three gated InAlAs/InGaAs/InAlAs
based quantum wells, which were designed to have different values
of Rashba SOI strength\cite{Koga}. At $V_{gate}=0 V$, the
characteristic spin-split energy $\Delta$ for the different
samples are $\Delta \approx 0.5, 1.5$ and $1.8$ meV. We will refer
to these samples as to samples 1, 2, and 3, respectively (see also
\cite{meijer}). A 14T magnet is used to generate $B_{\parallel}$
(i.e. Zeeman coupling), and split coils mounted on the sample
holder are used to independently control $B_{\perp}$. The electron
density and mobility at $V_{gate}=0 V$ are $n \simeq 7 \cdot
10^{15} m^{-2}$ and $\mu \simeq$ 4 $m^2/Vs$, respectively. All
measurements have been performed on $20$ x $80 ~\mu m$ Hall-bar
shaped devices, at T=1.6K.

Experimentally, the values of $\tau_{\phi}(B_{\parallel})$, which
quantify the TRS breaking at different values of $E_Z/E_{SOI}$,
are obtained from the magnetoconductance as a function of
$B_{\perp}$, at different fixed values of $B_{\parallel}$ (i.e.
$E_Z$). Specifically, from the quantitative analysis of the
line-shape of the resulting magnetoconductance curves
$\sigma(B_{\perp})$, the values of $\tau_{\phi}(B_{\parallel})$
can be extracted. For all details of extracting
$\tau_{\phi}(B_{\parallel})$ we refer the reader to \cite{meijer}.
Here we recall that we use the theory of Iordanskii, Lyanda-Geller
and Pikus (ILP) \cite{ILP}, in which $\tau_{\phi}(B_{\parallel})$
and the spin relaxation time $\tau_{s}(B_{\parallel})$ are the
only free parameters\cite{2D assumption}.

Figure 2 shows an example of the measured magnetoconductance
curves $\sigma(B_{\perp})$ at different values of $B_{\parallel}$
(open dots) for sample 3. The continuous lines superimposed on the
data represent the best fit to the ILP theory. We find very good
agreement between data and theory for all values of
$B_{\parallel}$, or equivalently, for all values of the ratio
$E_Z/E_{SOI}$. This kind of analysis has been performed for all
samples, and for different values of the electron density, elastic
scattering time and SOI strength\cite{el-el effects}.

Note that for $E_Z/E_{SOI} \geq 1$ the weak-antilocalization is
fully suppressed (see Fig. 2). Therefore, in the limit that
$E_Z/E_{SOI} \geq 1$, $\tau_{\phi}(B_{\parallel})$ is the only
free parameter in the ILP model to fit the data, and can be
determined with great accuracy. Only in the narrow region where
$E_Z \approx E_{SOI}$, the value of $\tau_{\phi}(B_{\parallel})$
is possibly determined with somewhat less accuracy, due to
potential $B_{\parallel}$-induced anisotropies in the spin
relaxation time.

\begin{figure}[t]
\begin{center}\leavevmode
\includegraphics[width=0.95\linewidth]{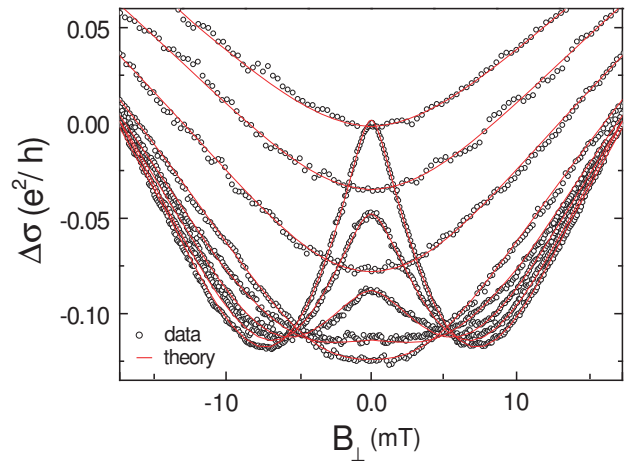}
\caption{(Color online) The magnetoconductance $\sigma(B_{\perp})$
of sample 3, measured at different values of $E_Z/E_{SOI}$: $0,
0.23, 0.46, 0.70, 1.16, 2.33, 3.72, \textrm{and }  4.66$,
corresponding to $B_{\parallel}= 0, 0.5, 1, 1.5, 2.5, 5, 8,
\textrm{and } 10$T. The solid lines represent the best fits to the
ILP theory, from which we obtain $\tau_{\phi}(B_{\parallel})$ and
$\tau_{s}(B_{\parallel})$, and hence also $E_{SOI} \equiv
\hbar/\tau_s(0)$.} \label{A}
\end{center}
\end{figure}

\begin{figure}[t]
\begin{center}\leavevmode
\includegraphics[width=0.95\linewidth]{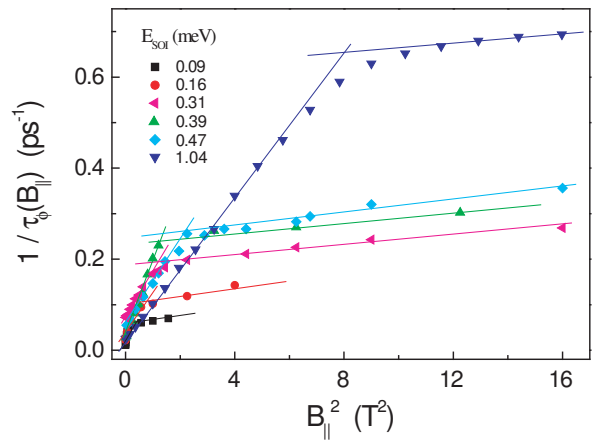}
\caption{(Color online) Extracted values of the dephasing rate
$1/\tau_{\phi}(B_{\parallel})$ as a function of $B_\parallel^2$.
The different symbols correspond to different values of $E_{SOI}$
\cite{nitta} (results from all three samples are shown). The
solid lines act as a guide to the eye.} \label{BB}
\end{center}
\end{figure}

In Figure 3 we first plot the extracted values of
$\tau_{\phi}(B_{\parallel})$ as a function of
$B_{\parallel}$ (or more precise, $B_{\parallel}^2$), since this
is the experimentally applied parameter. For each value of
$E_{SOI}$, we find the same qualitative behavior of the
$\tau_{\phi}(B_{\parallel})$-curve (results from all three
samples are shown). For small values of $B_{\parallel}^2$, the
slope $\partial (1/\tau_{\phi}(B_{\parallel}))/
\partial B_{\parallel}^2$ (hereafter called "dephasing-slope") is
large, and depends strongly on the value of $E_{SOI}$. For large
values of $B_{\parallel}^2$ (or $E_Z^2$) the dephasing-slope is
found to be much smaller. In both limits, we find that the
dephasing-slope is constant, i.e. $1/\tau_{\phi}(B_{\parallel})
\propto B_{\parallel}^2$. The value of $B_{\parallel}^2$ (or
$E_Z^2$) at which the crossover occurs is larger for larger values
of $E_{SOI}$. Anticipating, the crossover occurs when $E_Z/E_{SOI}
\approx 1$ (see Fig. 4).

The large dephasing-slope for $E_Z/E_{SOI} \ll 1$ is due to the
competition between Zeeman coupling and Rashba SOI\cite{meijer}.
In contrast, the small dephasing-slope in the high field limit can
be attributed to the coupling of $B_{\parallel}$ to the
\textit{orbital motion} of the electrons. Hence, Fig. 3 suggests
that the spin-induced dephasing of time reversed waves - i.e. the
spin-induced TRS breaking - \textit{saturates}.

To show the spin-induced part of the measured dephasing rate
$1/\tau_\phi(B_\parallel)$, we subtract the contribution due to
inelastic scattering ($\equiv 1/\tau_\phi(0)$), and denote the
spin-induced dephasing rate by $\Gamma^s_\phi(B_\parallel)$, with
$\Gamma^s_\phi(B_\parallel) = 1/\tau_\phi(B_\parallel) -
1/\tau_\phi(0)$. For $E_Z/E_{SOI} \ll 1$ the spin-induced
dephasing rate of time-reversed waves $\Gamma^s_\phi(B_\parallel)$
is given by\cite{malshukov} $\tau_s(0) \Gamma^s_\phi(B_\parallel)
= (E_Z/E_{SOI})^2$. In figure 4 we plot
$\tau_s(0)\Gamma^s_\phi(B_\parallel)$ for the whole measured range
of $E_Z/E_{SOI}$ \cite{g-factor}. For all samples, and all
different values of electron density, elastic scattering time,
Rashba strength, and $E_{SOI}$, the quantity
$\tau_s(0)\Gamma^s_\phi(B_\parallel)$ collapses to nearly the same
curve (the combined error in the determination of $\tau_s(0)$ and
$\Gamma^s_\phi(B_\parallel)$ is typically 10\%). We therefore
conclude that the spin-induced TRS breaking in quantum wells with
Rashba SOI (or more precisely $\tau_s(0)
\Gamma^s_\phi(B_\parallel)$) is a \textit{universal function} of
$E_Z/E_{SOI}$, within the experimental accuracy.

The spin-induced TRS breaking (or $\Gamma^s_\phi(B_\parallel)$)
saturates when $E_Z/E_{SOI} \approx 1$. For this strength of the
Zeeman coupling the spins start becoming parallel or anti-
parallel with $B_{\parallel}$. This conclusion is consistent with
the observation that for $E_Z/E_{SOI} \approx 1$ the
weak-antilocalization is just fully suppressed (see Fig. 2): when
the spins become aligned with $B_{\parallel}$, only
weak-localization is expected, since conceptually this situation
is identical to the case where only a small perpendicular field is
present, and the spins are aligned with $B_{\perp}$.

Currently, there are no theoretical predictions for the behavior
of $\tau_s(0) \Gamma^s_\phi(B_\parallel)$ when $E_Z/E_{SOI}$ is
not small, i.e. when the Zeeman coupling is not a small
perturbation. It has only been predicted, for specific cases, that
the magnetoconductance $\sigma (B_{\parallel})$ saturates when
$E_Z/E_{SOI} \gg 1$, indicating a saturation of $\tau_s(0)
\Gamma^s_\phi(B_\parallel)$\cite{malshukov2}. However, the
corresponding behavior of $\tau_s(0) \Gamma^s_\phi(B_\parallel)$ -
in particular its universal character - had not been recognized so
far.

\begin{figure}[t]
\begin{center}\leavevmode
\includegraphics[width=0.95\linewidth]{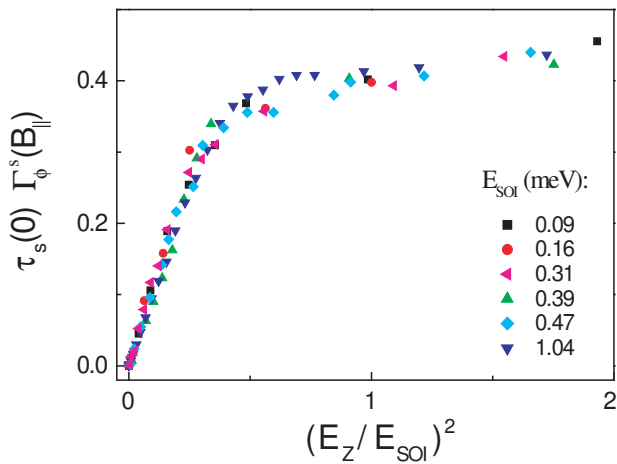}
\caption{(Color online) The spin-induced dephasing rate
$\Gamma^s_\phi(B_\parallel)$ multiplied by $\tau_s(0)$, as a
function of $(E_Z/ E_{SOI})^2$, with $\Gamma^s_\phi(B_\parallel) =
1/\tau_{\phi}(B_{\parallel}) - 1/\tau_{\phi}(0)$, and
$\tau_{\phi}(0)$ is the inelastic scattering time. Results from
all three samples are shown, illustrating that irrespective of
sample, electron density, SOI strength, etc., the data collapse on
nearly a single curve.} \label{C}
\end{center}
\end{figure}

We understand the saturation of $\Gamma^s_\phi(B_\parallel)$ for
$E_Z/E_{SOI} \gg 1$, and the dependence of the saturation value of
$\Gamma^s_\phi(B_\parallel)$ on spin relaxation time $\tau_s(0)$,
in the following intuitive way. Imagine first that the Zeeman
splitting is large and SOI is absent. At the Fermi energy, the
spin-split subbands are then well separated in k-space, and are
fully decoupled. In that case each subband contributes separately
to the interference, and the upper time-scale for interference is
the inelastic scattering time, independent of the size of the
Zeeman splitting. In the presence of SOI, the spin subbands become
weakly coupled, i.e. a particle can be scattered from one spin
subband to the other (flip its spin). Imagine there is a spin-flip
center at a certain position along the path. Both time-reversed
waves will then flip their spin at that position, and hence at
different times in general. The waves spend therefore different
amounts of time in each spin subband before they interfere, and
obtain a large phase difference, since
$k_{F,\uparrow}-k_{F,\downarrow}$ is large (large Zeeman
splitting). This implies that waves do no longer contribute to the
interference (on average) if a spin-flip event takes place along
the path. The upper time-scale for quantum interference is
therefore reduced to (roughly) the spin relaxation time
$\tau_s(0)$, independent of the Zeeman splitting, as long as
$k_{F,\uparrow}-k_{F,\downarrow}$ is large enough. This simple
picture is in qualitative agreement with our experiments.

Finally, we focus in more detail on the remaining small, but
finite, dephasing-slope for $E_Z/E_{SOI} \gg 1$ (see Figs. 3 and
4). To show the remaining dephasing rate for $E_Z/E_{SOI} \gg 1$
most clearly, we subtract the spin-induced dephasing rate in this
limit - which equals the saturation value
$1/\tau_{\phi}(E_Z/E_{SOI}=1)$ - and denote the resulting
dephasing rate by $\Gamma^{orb}_\phi(B_\parallel)$. Fig. 5 shows
the extracted values of $\Gamma^{orb}_\phi(B_\parallel) =
1/\tau_{\phi}(B_{\parallel})- 1/\tau_{\phi}(E_Z/E_{SOI}=1)$ for
sample 1 (weakest SOI), for different values of the electron
density\cite{high field} (Note that the $B^2_{\parallel}$-field
scale in this graph is much larger than in Figs. 3 and 4, i.e.
$E_Z/E_{SOI} \gg 1$). We find that the dephasing slope is larger
for larger values of the electron density. In particular, we find
that the remaining dephasing slope scales about linearly with the
Fermi velocity (see inset).

The finite thickness of the quantum well makes that
$B_{\parallel}$ does not only couple to the electron spin (Zeeman
coupling), but also to its \textit{orbital motion}. It has been
shown that this orbital coupling can also break TRS, via various
mechanisms\cite{mathur,TRS}. These mechanisms depend on the
specific (non-universal) details of the quantum well, such as
surface roughness, $z$-dependence of the scattering potential in
the 2DEG, and the asymmetry of the confining potential. The linear
dependence of $\Gamma^{orb}_\phi(B_\parallel)$ (or
$1/\tau_{\phi}(B_{\parallel})$) on $B_{\parallel}^2$, together
with the linear dependence of the dephasing-slope on $v_F$,
indicates that the (small) remaining slope for $E_Z/E_{SOI} \gg 1$
is caused by surface roughness\cite{mathur}. The resulting
dephasing-slope is given by $\partial
(1/\tau_{\phi}(B_{\parallel}))/
\partial B_{\parallel}^2 \approx \sqrt\pi e^2 d^2 L v_F/\hbar^2$,
where $d$ is the mean roughness height and L is the correlation
length of the roughness. For our quantum well we find $d^2 L
\approx 0.4 ~ nm^3$, which is a small value in comparison to other
reports\cite{minkov}.

In general, the orbital mechanism will break the universality of
the experimentally measured TRS breaking in systems with SOI,
since it adds to the (universal) spin-induced TRS breaking
mechanism. In our samples, the orbital TRS mechanism is very small
an hence unimportant, so it affects the universality only minorly.
This allows us to observe the universal behavior of the
spin-induced TRS breaking for all values of the ratio
$E_Z/E_{SOI}$.

In conclusion, we have demonstrated that the TRS breaking rate
$\Gamma_\phi(B_\parallel)$, caused by the competition between
Rashba and Zeeman, saturates when $E_Z \approx E_{SOI}$. This is
because for $E_Z \geq E_{SOI}$, two Fermi surfaces start being
formed with well-defined spin direction, pointing either parallel
or anti-parallel to $B_{\parallel}$. Moreover, we have shown that
the scaled dephasing rate, $\tau_s(0) \Gamma_\phi(B_\parallel)$,
is a universal function of the ratio $E_Z/E_{SOI}$, within the
experimental accuracy. Finally, we have shown that this
universality is broken by the coupling of the magnetic field to
the orbital motion of the electrons.

\begin{figure}[t]
\begin{center}\leavevmode
\includegraphics[width=0.9\linewidth]{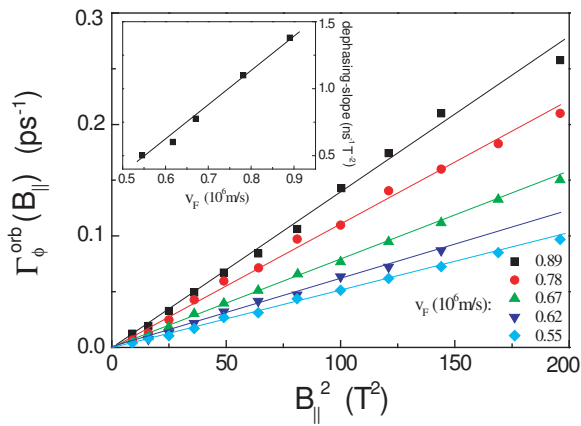}
\caption{(Color online) The measured values of
$\Gamma^{orb}_\phi(B_\parallel)=1/\tau_{\phi}(B_{\parallel})-1/\tau_{\phi}(E_Z/E_{SOI}=1)$
in sample 1, for various values of the electron density (i.e.
$v_F$). The dephasing-slope, $\partial
(1/\tau_{\phi}(B_{\parallel}))/
\partial B_{\parallel}^2 = \partial
\Gamma^{orb}_\phi(B_\parallel)/
\partial B_{\parallel}^2$, depends about linearly on $v_F$ (see inset).
This indicates that surface roughness is the main orbital TRS
breaking mechanism. Note that the $B^2_\parallel$-scale is much
larger than in Figs. 3 and 4.} \label{D}
\end{center}
\end{figure}

We would like to thank T. Koga for the sample fabrication, and Y.
M. Blanter, Y. V. Nazarov, and H. Takayanagi for stimulating
discussion and support.


\begin{thebibliography}{99}
\bibitem{time scale} For a perpendicular field this upper
time-scale is often denoted $\tau_B$, and is given by $\tau_B=h/2
e B D$, with $D$ is the diffusion constant.
\bibitem{2D assumption} In the ILP model, the magnetoconductance is assumed to be
determined by 2D interference, i.e $\sqrt{D
\tau_{\phi}(B_{\parallel})/3}$ must be much larger than the sample
thickness. This condition is satisfied for all values of
$B_{\parallel}$ in our measurements.
\bibitem{el-el effects} Notice that due to the relatively large electron
density and small effective mass ($m^* \approx 0.041 m_0$) the
electron gas is far from being spin-polarized, even at
$B_{\parallel}=14$T. Also, due to the large SOI strength in our
samples, together with $r_s \ll 1$, the magnetoconductance is
essentially only determined by interference corrections, and not
by $B_{\parallel}$-induced electron-electron interaction
effects\cite{el-el}. We do find a small quadratic enhancement of
the Drude resistivity with $B_{\parallel}$ ($ < 4\%$ at
$B_{\parallel}=14$T), the discussion of which is beyond the scope
of the present paper.
\bibitem{g-factor} The g-factor is obtained experimentally from
the TRS breaking for $E_Z/E_{SOI} \ll 1$, as shown in
\cite{meijer}.
\bibitem{high field} In order to obtain the highest possible accuracy
in determining $\tau_{\phi}(B_{\parallel})$, as it is needed to
determine the weak $B_{\parallel}$-dependence for $E_Z/E_{SOI} \gg
1$, we have (also) used the theory of Kawabata and
Dyakonov\cite{dyakonov} to extract $\tau_{\phi}(B_{\parallel})$.
Note that this theory can not be used for $E_Z/E_{SOI} \ll 1$, as
weak-antilocalization due to Rashba SOI is neglected.
\end{thebibliography}
\end{document}